\documentstyle[12pt,aasms]{article}

\def\etal{{\it et~al.\ }}

\tightenlines

\received{April 9, 1993}
\accepted{July 7, 1993}

\begin{document}

\title{Counter-Evolution of Faint Field Galaxies\altaffilmark{1}}

\author{David C. Koo\altaffilmark{2}}
\affil{University of California Observatories / Lick Observatory, \\
Board of Studies in Astronomy and Astrophysics, \\
University of California, Santa Cruz, California 95064}

\author{Caryl Gronwall}
\affil{Board of Studies in Astronomy and Astrophysics, \\
University of California, Santa Cruz, California 95064}

\author{Gustavo Bruzual A.} 
\affil{Centro de Investigaciones de Astronomia (CIDA) \\
A.P. 264, M\'{e}rida, Venezuela}

\hskip 1.0in
\centerline{\it To be published in the Astrophysical Journal Letters}

\altaffiltext{1}{UCO / Lick Observatory Bulletin No. 1263}
\altaffiltext{2}{Visiting Astronomer, Kitt Peak National Observatory,
National Optical Astronomy Observatories, which is operated by 
the Association of Universities for Research in Astronomy, Inc. (AURA)
under cooperative agreement with the National Science
Foundation.} 

\dates

\newpage
\begin{center}
	\large\bf Abstract
\end{center}

We adopt a new approach to explore the puzzling nature of faint blue 
field galaxies. 
Instead of assuming that the local luminosity function is well
defined, we first determine whether {\it any} non-evolving
set of luminosity functions for different spectral types of galaxies is
compatible with the observed marginal distributions in optical and 
near-infrared counts, $B - R$ colors, and redshifts. Exploiting a 
non-negative
least squares method, we derive a new  no-evolution model that is
found to fit {\it all}
the observations surprisingly well, from $B \le 15$ to the limit of 
$B \sim 24$ for redshifts and
$B \sim 26$ for colors and counts. Contrary to previous conclusions,
the faint galaxies in excess of our new non-evolutionary model are red
rather than blue.  Although our fits are far better than previous
no-evolution models, the remaining deviations from the observations
still suggest the need for some, but slight,
evolutionary component or model revisions.
We conclude that 
models more exotic than mild luminosity evolution, such as those 
requiring rapid evolution in star formation rates, disappearing dwarf galaxy
populations, high values of the cosmological
constant, rapid mergers, or substantial 
non-conservation of galaxy numbers with time,
are no longer as compelling. Our results resurrect field galaxies as
promising probes of the curvature of space.



\clearpage
\section{Introduction}

For over a decade, the steep slope of the faint field galaxy counts
and their very blue colors have been explained as the result of mild 
evolution in the spectral energy distributions (SED's) of galaxies. 
The excess of counts above non-evolutionary (NE) models has been
claimed to be a factor of two  by $B \sim 20$ 
(Maddox \etal 1990, Loveday \etal 1992) or by  
$B \sim 22.5$ (Colless \etal 1990),
and  even factors of 5 to 15 by $B \sim 25$ (Tyson 1988). 
These blue counts have been claimed, even with evolution, to be incompatible
with a flat ($\Omega = 1$) Friedmann universe (Koo 1990, Guiderdoni and 
Rocca-Volmerange 1990)
or with the observed near-infrared K band counts (Cowie \etal 1993).
Yet recent redshift surveys show
that galaxies fainter than 20th mag exhibit redshift distributions close
to that predicted by NE models (Broadhurst \etal 1988, Colless \etal 1990,
Lilly \etal 1991). 

A commonly held belief is that dramatic revisions to
the conventional view are needed. Some have revised the cosmology by
adopting a non-zero cosmological constant, $\Lambda$ (Fukugita \etal
1990).  Others
have proposed non-conservation of galaxy numbers, due to mergers 
(White 1989, Cowie \etal 1991, Broadhurst \etal 1992), a disappearing 
population
of dwarf galaxies  (Cowie \etal 1991, Babul and Rees 1992), 
or dwarfs which have faded substantially
in recent times (Broadhurst \etal 1988).  

An alternative  view is that the uncertainties of local and distant
field galaxy data and galaxy models preclude a convincing case
for any exotic theories at this time.
Instead, a  NE model generated by
{\it trial and error} adjustments of galaxy luminosity functions (LF's)
versus color is claimed to provide moderately good fits to the
known data (Koo and Kron 1992). In this letter, we adopt a new 
{\it objective} method to
answer the question: what is the best fit that {\it any} NE model 
can ever hope to 
make to the observations?

\section{Observations and Method}

Our method of fitting a model to the data is conceptually simple.  
The goal is to solve for the weights to be applied to a set of input
basis vectors that best fit (using non-negative least squares) the
observations. Our technique is similar to that used to derive the
stellar components of a composite galaxy spectrum in population
synthesis studies (e.g. Faber 1972).
Rather than the spectra for stars used in population synthesis,
the input vectors in our current analysis are the
predicted number distributions of colors and redshifts for 154
combinations of galaxy absolute luminosities and color-classes.
Fourteen absolute magnitude bins were chosen, each extending from
a bright limit to a faint limit of $M_{B_J}$ = -11 (i.e., the first
bin is one
magnitude wide extending from $M_{B_J}$ = -12 to -11, the next is two
magnitudes wide extending from $M_{B_J}$ = -13 to -11, and so on).
Bins were chosen in this way to ensure
that the different LF's were both monotonically increasing and smooth
(dropping these constraints did not significantly improve 
the fits).  
An open universe with $q_o =
0.05$ and a Hubble constant of 50 km-sec$^{-1}$-Mpc$^{-1}$
is adopted for the model predictions.  Eleven color classes of
galaxies were chosen, ranging from very red to very blue (see Table
1).  Recently revised Bruzual models (Bruzual and Charlot 1993) were
used to create the SED's.  For NE, the SED's, i.e.  luminosities and
colors, of the galaxies are assumed to be independent of time.  The
output weights for these 154 kinds of galaxies are proportional to
their volume densities, and thus provide a direct measure of the LF
for each color-class of galaxy. 

Instead of the galaxy spectrum used in population synthesis, 
the data to be fit 
in this study is organized as a vector matched to the model input
vectors and consists of the observed galaxy counts (in $K$, $r$, and
$B_{\rm J}$) as well as
color and redshift distributions. The color data
were binned in 0.1 mag of $B_{\rm J} - R_{\rm F}$ and ranged from $B
\sim 19$ to $27$ at one magnitude intervals. The redshift data ranged
from $B \sim 15$ to $24$, also at one magnitude intervals, but binned
in 1.0  units of 7 log z.  Each color or redshift 
distribution was {\it normalized} to
the observed $B_{\rm J}$ counts.  These observations, along with their
attendant problems of zero-points for the colors and incompleteness
for the redshifts, are from the compilations of Koo and Kron (1992),
except for 
the recent redshifts of Colless
\etal (1993) for $B = 23$ to $24$.  Though inclusion of observational
errors are important at some level, they were largely ignored.
The results also depend on the relative 
weights adopted for
different parts  of the data set. As a default, we have used weights
which are proportional to the observed number of galaxies in each bin.
However, in order to de-emphasize the importance of the number counts we
fit to the number of galaxies per 0.1 square degrees instead of per 1
square degree.  In addition, we have weighted the redshift
distributions by 5 times the observed numbers in each bin to increase
the importance of the observed redshift distributions relative to the
color distributions.

In mathematical terms, given an
$ m \times n$ matrix, $P$, and an $m$-vector, $D$, we wish to compute the
$n$-vector, $X$, which solves the least squares problem,
PX \ =\ D
, subject to $X_i \geq 0$ for all $i$.
The data vector, $D$, is filled by the tables of observational data.  
The $ith$-column of
the prediction matrix, $P$, is an input vector filled by the
model-generated table of data for each kind of galaxy, $i$. 
The output vector, $X_i$, represents the number per unit volume 
assigned to the
$ith$ kind of galaxy in the least-squares solution.

\section{Results and Discussion}

Table 2 provides the resultant LF for each of the Bruzual color
classes. Figure 1 compares various
color-integrated LF's to some other recent derivations.  Our total LF
to $M_{B_J} \sim -18$ is seen to be a good match to the flat local LF
derived by Loveday \etal (1992). Though our predicted rise appears
inconsistent with the faintest two points of the Loveday \etal LF,
Figure 1 shows that this steepening is compatible with recent LF's
derived from fainter redshift surveys, including the {\it local} LF
derived by Eales (1993) and the faint field ($B_J > 20$) LF derived by
Lonsdale and Chokshi (1993).  Our LF's divided by color can be
directly compared to those adopted by Metcalfe \etal (1991); our LF
when summed over the same color intervals of $B-V$ (see Figure 1) also
show a steep
low-luminosity rise for the blue galaxies and consistent
normalizations brighter than the valid limits of the Metcalfe \etal
(1991) fits.

Using these LF's,
the predicted NE counts, colors, and redshift distributions are
compared to existing observations and displayed in Figures 2 to 4,
respectively. The fits are remarkably good and considerably improved
over the hand-made NE model of Koo and Kron (1992). Though not shown,
we also checked that the color-redshift distributions are also
consistent, since in principle, different correlations of
color-redshift may result in the same color or redshift marginal
distribution.  Our new NE model shows
only a slight deficit (less than a factor of two) of blue counts
compared to the observations,
even to the faintest reliabele limits of $B \sim 25$.  The fit to the
red ($r$)
counts is excellent and the fit to the near-infrared ($K$) counts is
acceptable.  The spectral energy distributions predict $V-K$ colors
for elliptical galaxies that are too blue by approximately 0.3 mag
compared to observations (see Fig. 11, Bruzual and Charlot 1993).  A
shift of 0.3 mag towards brighter magnitudes in K for the fit to the
K-band counts would actually improve the fit.  The color distributions
are also very well matched from $B \sim 19$ to $25$. It is intriguing
to note that the comparison of the color distributions to the NE model
shows that the {\it excess faint galaxies are redder than average, not
bluer}, contrary to standard lore based on previous NE models.
The match to the
redshift distributions is generally not as good fainter than about
21st mag, beyond which the predictions yield more low redshift $z <
0.1$ (i.e. local low-luminosity dwarfs with $M_{\rm B} \sim -12$ to
$-17$) galaxies than observed, typically by about 10\% of the total
sample.  Whether this slight discrepancy can be explained by selection
effects against compact or low surface-brightness dwarfs in the
observations, by systematic errors in the zero-points of the faintest
color data, or by the limitations of our model remains to be
explored. For $B = 23$ to $24$, our new NE model predicts that about  
33\% of
the galaxies will have redshifts $z \ge 0.7$. This is greater than the 15\%
observed in the most recent data of Colless \etal (1993), but not
at a high level of statistical significance given the sparse redshift
sample at these very faint magnitudes.

Despite the success of our near-optimal NE model, 
we are not arguing against some 
evolution. To the extent that the observed data are all accurate and
our SED's are representative of real galaxies, 
our experiment indicates that
no NE model exists that  will
fit the observations totally, especially the 
higher (by $\sim 40\%$) counts by $B_J \sim 24$. 
Whether these discrepancies can be explained by improved data and/or
various evolutionary models will be explored in future papers along
with more detailed error analysis and multicolor data.

Since our NE model is able to fit the counts, colors, and redshifts so
well over a large range in magnitude, we explored why previous models
predict fewer faint galaxies and redder colors.  Most assumed
one-to-one conversions of galaxy morphology to color 
rather than including a dispersion of colors for each galaxy type; 
this assumption
naturally results in too many red galaxies compared to reality. 
Others adopted a single LF
shape for all galaxy types; with a standard Schechter LF, e.g., the
blue galaxies tend to be underestimated at the extremes of the LF.
Many models of the bluest galaxies included a characteristic
luminosity for either a Schechter or Gaussian LF that was fainter than
that for redder galaxies; this tends to underestimate the number of
luminous blue galaxies, which are known to exist locally (specific
examples can be found in the study of
Markarian galaxies by  Huchra 1977).  
Some models even had the bluest galaxies as those
with constant star-formation rate and thus $B-V \sim 0.45 - 0.55$;
many galaxies are bluer (again, see Huchra 1977). 

In conclusion, we confirm our contrarian view from Koo and Kron (1992)
that exotic theories are not yet 
required to explain existing faint field galaxy
data.  Instead, we suggest that the local LF's for different color
classes of galaxies adopted by previous studies are likely to be
significantly in error.  We derive a plausible set of LF's by an
objective technique.  The resulting NE predictions match well enough
to the observations so that adoption of mild luminosity evolution
remains a viable path to improved fits.  
Since the observational evidence for non-conservation of
galaxy numbers with recent lookback time is now far less compelling,
distant field galaxies may be resurrected as promising probes of the
curvature of space.

\acknowledgments

We thank M. Bolte, J. Huchra, 
G. Illingworth, R. Kron, 
J. Ostriker, and the referee for useful discussions and suggestions
that improved the
clarity of the paper.  We gratefully acknowledge support for this work
from an US-Venezuela NSF grant INT-9003157, an NSF PYI grant
AST-8858203, two faculty research grants from the University of
California, Santa Cruz, and a Bilateral International Cooperation
Project (PI-027) from CONICIT-Venezuela. C. G. acknowledges partial
support from an NSF Graduate Fellowship.

\def\kms{{km~s$^{-1}$}}

\clearpage

\begin{table*}
\centerline {TABLE 1}
\centerline {INPUT GALAXY MODELS}
\begin{center}
\begin{tabular}{crrrrrrrrrrr}
\tableline
\tableline
Model    &1 &2 &3 &4 &5 &6 &7 &8 &9 &10&11\\
\tableline
Age\tablenotemark{(a)} &0.4    &2.0    &6.8    &16.0    &16.0    &16.0
&16.0   &16.0   &16.0   &16.0   &20.0\\
$\mu$\tablenotemark{(b)} &C&  C&  C &
0.01&   0.10& 0.15&   0.20&    0.25&     0.30&     0.70&  B\\
$B-V$  &0.15    &0.25    &0.35    &0.43    &0.52
&0.61    &0.70    &0.78    &0.85    &0.95   &0.99\\
$B_{J}-R_{F}$    & 0.26   & 0.38   & 0.49   & 0.59  & 0.69  &  0.78
& 0.87  &  0.96    &1.02   &1.13    &1.17\\
\tableline
\end{tabular}

\tablenotetext{a}{ Age of the galaxy model in Gyr.}
\tablenotetext{b}{ Star formation history of the galaxy model:  All
models have Salpeter initial mass functions for masses between 0.1 and
125 $M_\odot$.  Models 1-3 have a constant (C)  star formation rate.
Models 4-10 have an exponentially decreasing star formation rate
parameterized by $\mu$, where $\mu$ is the fraction of mass converted
into stars in 1 Gyr.  Model 11 is a $10^7$ Gyr burst (B)  of star formation.}
\end{center}

\end{table*}
\clearpage
\begin{table*}
\centerline{TABLE 2}
\centerline{OUTPUT MODEL LUMINOSITY FUNCTIONS}
\centerline{LOG $\Phi$ (Number/mag/Mpc$^3$)\tablenotemark{(a)}}
\begin{center}
\begin{tabular}{crrrrrrrrrrrrr}
\tableline
\tableline
   &\multicolumn{10}{c}{Model}&  &  \\
$M_{B_{J}}\tablenotemark{(b)}$ &1 &2 &3 &4 &5 &6 &7 &8 &9
&10&11&Total&Obs.\tablenotemark{(c)}\\
\tableline
-24.5&     &     &     &     &     &     &     &-8.72&     &      
&     &-8.72&-10.8\\
-23.5&-6.66&     &     &     &-7.05&-7.07&-6.32&-8.72&     &      
&     &-6.06&-6.32\\
-22.5&-6.66&     &     &     &-7.05&-4.74&-5.04&-8.72&     &      
&     &-4.56&-4.35\\
-21.5&-5.40&-4.64&-4.51&-4.12&-7.05&-3.93&-4.07&-4.30&     &      
&     &-3.41&-3.46\\
-20.5&-3.75&-4.64&-3.97&-3.37&-4.34&-3.93&-4.07&-4.30&     &      
&-3.67&-2.90&-3.07\\
-19.5&-3.59&-4.64&-3.97&-3.37&-3.49&-3.93&-4.01&-4.06&     &      
&-3.34&-2.72&-2.92\\
-18.5&-3.59&-4.64&-3.97&-3.37&-3.49&-3.27&-4.01&-4.06&     &      
&-2.99&-2.54&-2.87\\
-17.5&-3.59&-4.64&-3.97&-3.37&-3.49&-3.27&-4.01&-4.06&     &      
&-2.84&-2.48&-2.85\\
-16.5&-3.59&-4.64&-3.97&-3.37&-3.49&-3.27&-4.01&-4.06&     &      
&-2.32&-2.18&-2.85\tablenotemark{(d)}\\
-15.5&-3.59&-4.64&-3.97&-3.37&-3.49&-3.27&-4.01&-4.06&     &      
&-2.04&-1.96&-2.86\\
-14.5&-2.24&-4.64&-3.97&-1.75&-3.49&-3.27&-4.01&-4.06&     &      
&-2.04&-1.47&-2.87\\
-13.5&-2.24&-4.64&-1.14&-1.75&-3.49&-3.27&-4.01&-4.06&     &      
&-2.04&-0.97&-2.88\\ 
-12.5&-2.24&-4.64&-0.62&-1.75&-3.49&-3.27&-4.01&-4.06&     &      
&-2.04&-0.57&-2.89\\
-11.5&-2.24&-4.64&-0.10&-1.75&-3.49&-3.27&-4.01&-4.06&     &      
&-2.04&-0.08&-2.91\\  
\tableline
\end{tabular}
\tablenotetext{a}{ H$_o$=50 \kms-Mpc$^{-1}$}
\tablenotetext{b}{ Center of one magnitude bins}
\tablenotetext{c}{ Luminosity function from Loveday \etal (1992)):
$M^{*}$ = -21, $\alpha = -0.97$}
\tablenotetext{d}{ Loveday \etal luminosity function only valid for
$M_{B_J}$ $\leq -16.75$}
\end{center}



%

\end{table*}

\clearpage
\title {Figure Captions}

{\sc fig}. 1a. -- {Derived luminosity functions  versus previously
derived 
ones, all scaled to 
 H$_o = 50$ km-sec$^{-1}$-Mpc$^{-1}$. Thick-lined histogram is the
derived total differential luminosity function; thin line is from the derived
Schechter LF of
Loveday \etal (19992) and is valid for $M_{B_J}~\leq~-16.75$ (dashed
line indicates extrapolation of this fit).  Circles are data
points from Loveday \etal (1992), plus signs from
Eales (1993), and stars from Lonsdale and Chokshi
(1993).  For clarity, we have not included the authors' original
error bars on these points.
}

{\sc fig}. 1b. -- {Same as for 1a, except thick-lined histogram applies for 
color classes with $B-V \geq 0.85$.  The thin line is the 
differential LF from Table 8 of Metcalfe \etal (1991) for
the same color range and valid for $M_{B_J} \leq -17.5$; dashed
line indicates extrapolation of their fit. 
}

{\sc fig}. 1c. --  {Same as for 1b except for 
$0.6 < B-V < 0.85$ and thin line being valid for $M_{B_J} \leq -15.5$; dashed
line indicates extrapolation of their  fit.
}

{\sc fig}. 1d. --  {Same as for 1c except for 
$B-V \leq 0.6$. 
}

{\sc fig}. 2. -- {Log of the differential counts A (per mag per square degree)
of faint field galaxies versus magnitude 
in the indicated bands. The $B_{\rm J}$ and
Gunn $r$ band observations (symbols)  are compilations of data made by 
Koo and Kron (1992); the infrared $K$ band counts are a compilation from
Gardner \etal (1993). Our new no-evolution predictions are shown as curves.
}

{\sc fig}. 3. -- {Color ($B_{\rm J} - R_{\rm F}$) distributions versus
indicated magnitude
intervals are shown as thick lines for the observations and thin lines
for the model
predictions. The observations are compilations (Koo and Kron 1992)
from the data of several groups.
}

{\sc fig}. 4. -- {Normalized histograms versus redshift (log z)  
for  magnitude ($B_{\rm J}$)
intervals as indicated.
Bin size is 0.143, since original bin size was 1.0 in 7log z.
The model predictions are shown as thin lines; 
observations shown as thick lines are compilations from different
surveys (Koo and Kron 1992); the $B_{\rm J} = 23$ to $24$ observations
are updated from Colless~\etal~(1993).  
}

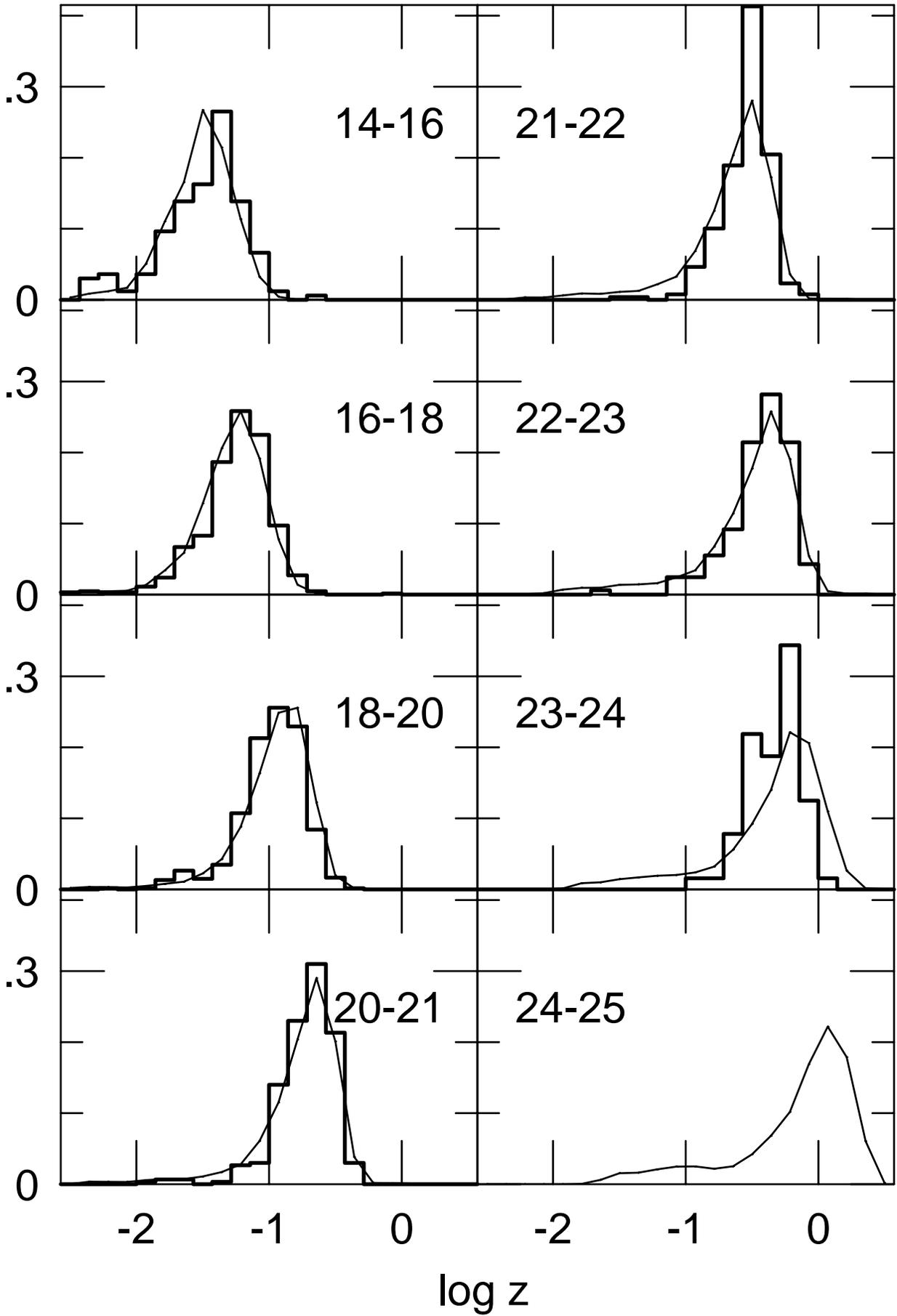
\end{document}